\documentclass[fleqn,usenatbib,usedcolumn]{mnras}

\usepackage[british]{babel}	
\usepackage{txfonts}	

\usepackage[T1]{fontenc}	
\usepackage{ae,aecompl}

\usepackage{graphicx}	

\newcommand{\abs}[1]{\mbox{$\left|{#1}\right|$}}

\newcommand{\fracd}[2]{\frac{\displaystyle{#1}}{\displaystyle{#2}}}
\newcommand{\dd}{{\rm d}}
\newcommand{\ricci}{\mbox{$\cal R$}}
\newcommand{\acg}{\mbox{$\alpha_g$}}
\newcommand{\muf}{\mbox{$\mu$}}
\newcommand{\lag}{\mbox{$\cal L$}}
\newcommand{\bs}{\mbox{$\!\!\!$}}
\newcommand{\psibar}{\mbox{$\bar{\psi}$}}
\newcommand{\msun}{\mbox{M$_\odot$}}
\newcommand{\dslash}{\mbox{$\,\not\bs{\cal D}$}}

\title[ Conformal Gravity Rotation Curves ]
{ Conformal Gravity Rotation Curves with a Conformal Higgs Halo }

\author[Horne]
{Keith Horne$^{1}$\thanks{E-mail: kdh1@st-and.ac.uk}
\\ $^{1}$SUPA Physics and Astronomy,
	University of St.~Andrews,
	KY16 9SS, Scotland, UK
}

\date{Accepted 2016 February 28. Received 2016 February 25; in original form 2016 January 27}

\pubyear{2016}

\begin{document}
\label{firstpage}
\pagerange{\pageref{firstpage}--\pageref{lastpage}}
\maketitle


\begin{abstract} 
We discuss the effect of a conformally coupled Higgs field on 
Conformal Gravity (CG) predictions for the
rotation curves of galaxies.
The Mannheim-Kazanas (MK) metric is a valid vacuum solution
of CG's 4-th order Poisson equation if and only if the
Higgs field has a particular radial profile,
$S(r)=S_0\,a/(r+a)$, decreasing from $S_0$ 
at $r=0$ with radial scale length $a$.
Since particle rest masses scale with $S(r)/S_0$,
their world lines do not follow time-like geodesics
of the MK metric $g_{\mu\nu}$, as previously assumed, but
rather those of the Higgs-frame MK metric 
$\tilde{g}_{\mu\nu}=\Omega^2\,g_{\mu\nu}$,
with the conformal factor $\Omega(r)=S(r)/S_0$.
We show that the required stretching of the MK metric 
exactly cancels the linear potential 
that has been invoked to fit galaxy rotation
curves without dark matter.
We also formulate,
for spherical structures with a Higgs halo $S(r)$,
the CG equations that must be solved for
viable astrophysical tests of CG using
galaxy and cluster dynamics and lensing.
\end{abstract}





\section{Introduction}
\label{sec:intro}
\typeout{ Introduction }

The need for dark matter and dark energy
to reconcile Einstein's General Relativity (GR)
with observations, together with the lack of
other tangible evidence for their existence,
motivates the study of alternative gravity theories
aiming to achieve similar success without
resort to the dark sector.

Conformal Gravity (CG), like GR, employs
a metric to describe gravity as curved space-time.
But the CG field equations, which dictate how
matter and energy generate space-time curvature,
arise from a local symmetry principle,
conformal symmetry, which holds for the strong,
weak and electro-magnetic interactions,
but is violated by GR.
Conformal symmetry means that stretching the metric
by a factor $\Omega^2(x)$, and scaling all other
fields by appropriate powers of $\Omega$, 
has no physical effects.
In particular, local conformal transformations preserve
all angles and the causal relations among
space-time events, but physical distances, time
intervals, and masses change, so that only local
ratios of these quantities have physical significance.
 
Unlike GR, and many related alternative gravity theories,
terms allowed in the CG action are highly restricted
by the required conformal symmetry.
GR's Einstein-Hilbert action adopts the Ricci scalar $\ricci$,
leading to Einstein's famous 2nd-order field equations,
\begin{equation}
	G_{\mu \nu} + \Lambda\,g_{\mu \nu}  = -8\,\pi\,G\,T_{\mu \nu}	
\ ,
\end{equation}
where $G_{\mu \nu} = R_{\mu \nu} - \left(\ricci/2\right)\,g_{\mu \nu}$
is the Einstein tensor, $G$ is Newton's constant,
and $\Lambda$ is the cosmological constant.
These terms are excluded in CG because $\Lambda$ and
$G$ build in fundamental scales, and $\ricci$ violates
conformal symmetry.
Instead the CG action allows only conformally-invariant
scalars linked by dimensionless coupling constants.
$\ricci$ can appear if it is coupled to a scalar field
$S$ in the particular conformally-invariant combination
$
	S^{;\mu}\,S_{;\mu} - \fracd{\ricci}{6}\,S^2
\ .
$
Particle rest masses cannot be fundamental, but may instead
arise through Yukawa couplings to the conformal
Higgs field $S$.
Likewise, the Higgs mass cannot be fundamental, but may
arise through dynamical symmetry breaking.
Conformal invariance replaces Einstein's 2nd-order field equations
with the 4th-order CG field equations \cite{mk89,m06},
\begin{equation}
        4\, \acg\, W_{\mu \nu} = T_{\mu \nu}
\ ,
\end{equation}
where $\acg$ is a dimensionless coupling constant.
The Bach tensor $W_{\mu\nu}$ and 
stress-energy tensor $T_{\mu\nu}$
are both traceless, and scale as $\Omega^{-4}$.

Despite their complexity, the 4th-order CG field equations
admit analytic solutions for systems with
sufficient symmetry \cite{m06}.
For homogeneous and isotropic space-times \cite{m01},
the Robertson-Walker metric is a solution with
$W_{\mu \nu}=0$, providing a dynamical cosmological model
identical to that of GR, except that the Friedmann equation
has a negative effective gravitiational
constant $G_{\rm eff}=-3/4\,\pi\,S_0^2$,
where $S_0$ is the vacuum expectation value of 
a conformally-coupled scalar field with
vacuum energy density $\lambda\,S_0^4$.
This CG cosmology gives an open universe,
but it can fit luminosity distances
from supernovae \cite{m03}
and features cosmic acceleration with 
$0<\Omega_\Lambda<1$,
neatly solving the cosmological
constant problem without dark energy
\cite{m01,m11}, see also \cite{n11}.
Growth of structure in CG cosmology is 
starting to be investigated \cite{m12},
but has not yet produced predictions for the CMB.

The plan of this paper is as follows:
In Sec.~\ref{sec:mk} we review the static spherical solutions
that have been used with some success \cite{m93,m97,mo12} to fit
the rotation curves of spiral galaxies, large and small.
In Sec.~\ref{sec:higgs} we discuss the need for
a conformally coupled Higgs field $S(r)$,
and show that
it makes a non-zero contribution to the source $f(r)$
in CG's 4th-order Poisson equation 
unless it has a particular radial
profile $S(r)=S_0\,a/(r+a)$.
In Sec.~\ref{sec:frame} we stretch the MK metric
with the conformal factor $\Omega_S(r)=S(r)/S_0$,
and show that the linear potential used in 
previous fits to galaxy rotation curves is
effectively removed.
In Sec.~\ref{sec:eqns} we summarise the
coupled system of equations that must be solved
in order to make astrophysical tests of CG predictions for
static spherically symmetric structures.
We summarise and conclude in Sec.~\ref{sec:fini}.

\section{Static Spherical Solutions}
\label{sec:mk}
\typeout{Static Spherical Solutions}

For static and spherically symmetric spacetime geometries,
analytic solutions to CG include 
the Mannheim-Kazanas metric \cite{mk89} (MK),
an extension of GR's Schwarzschild metric.

Co-moving coordinates render $W^\mu_\nu$ and $T^\mu_\nu$
diagonal, giving in principle 4 CG field equations.
But only 2 are independent, given that
spherical symmetry requires $W^\theta_\theta=W^\phi_\phi$,
and the Bianchi identities require a traceless
Bach tensor, $W^\mu_\mu=0$, and hence $T^\mu_\mu=0$.

MK show that for any static spherically symmetric spacetime,
a particular conformal transformation
brings the metric into a standard form
\footnote{Here and henceforth we adopt natural units, $\hbar=c=G=1$.},
\begin{equation}
	\dd s^2 = -B(r)\, \dd t^2
	+ \fracd{\dd r^2}{B(r)}
	+ r^2\, \dd \theta^2
	+ r^2\, \sin^2{\theta}\, \dd \phi^2
\ .
\end{equation}
We refer to this standard form, in which
$-g_{00}=g_{rr}=B(r)$, as the ``MK frame''.

With this metric ansatz, MK show that the CG field equations
boil down to an exact 4th-order Poisson equation for $B(r)$:
\begin{equation}
\label{eqn:poisson}
	\fracd{3}{B}
	\left( W^0_0 - W^r_r \right)
	= \fracd{1}{r}\,
	\left( r\,B \right)^{\prime\prime\prime\prime}
        = \fracd{ 3 }{4\,\acg\,B} \left( T^0_0 - T^r_r \right)
	\equiv f(r)
\ ,
\end{equation}
where $^\prime$ denotes $\dd/\dd r$,
and $f(r)$ is the CG source.
Note that $T^0_0=-\rho$ and $T^r_r=p$
for a perfect fluid (representing matter and radiation)
with energy density $\rho$ and pressure $p$.
We then require $\acg<0$ so that
a localised source with $\rho-p>0$ generates
attractive gravity.
\footnote{While \cite{f06} argue that CG is repulsive
in the Newtonian limit, \cite{m07} show that this
mistaken conclusion arises from a subtlety
in taking the Newtonian limit in isotropic coordinates,
and that $\acg<0$ gives locally attractive 
gravity in the Newtonian limit.
}

The remaining constraint can then be the 3rd-order equation
\begin{equation}
\begin{array}{rl}
	W^r_r
& \bs \bs
	= \fracd{1-B^2}{3\,r^4}
	+ \fracd{2\,B\,B^\prime}{3\,r^3}
	- \fracd{B\,B^{\prime\prime}+\left(B^\prime\right)^2}{\,3r^2}
\\ & \bs \bs
	+ \fracd{B^\prime\,B^{\prime\prime}-B\,B^{\prime\prime\prime}}{3\,r}
	+ \fracd{2\,B^\prime\,B^{\prime\prime\prime}
	- \left(B^{\prime\prime}\right)^2}{12}
	= \fracd{1}{4\,\acg}\,T^r_r
\ ,
\end{array}
\end{equation}
which can be imposed as a boundary condition \cite{bv09}.

\subsection{Vacuum Solution : The MK metric}
\typeout{Source-free Static Spherical Solutions}

A {\it source-free} vacuum solution requires
$T^0_0 = T^r_r$, so that $f(r)$=0.
Since
\begin{equation}
	\left( r^{n+1} \right)^{\prime\prime\prime\prime}
	= (n+1)\,n\,(n-1)\,(n-2)\,r^{n-3}
\ 
\end{equation}
vanishes for $n=-1$, 0, 1, and 2,
the homogeneous 4th-order Poisson equation then integrates 
4 times to give
\begin{equation}
        B(r) = w - \fracd{2\,\beta}{r} + \gamma\, r - \kappa\,r^2
\ ,
\end{equation}
with 4 integration constants $w$, $\beta$, $\gamma$ and $\kappa$.
 The 3rd-order constraint $4\,\acg\,W^r_r=T^r_r$ then gives
\begin{equation}
        w^2 = 1 - 6\,\beta\,\gamma
	+ \fracd{3\,r^4}{4\,\acg}\,T^r_r
\ .
\end{equation}
For $T^r_r=0$, or $T^r_r\propto r^{-4}$,
this gives 1 constraint on the 4 coefficients,
leaving the metric with 3 parameters:
$\beta$, $\gamma$, $\kappa$.

 This Mannheim-Kazanas (MK) metric matches the 
successes of GR's Schwarzschild metric
in the classic solar system tests, if we identify $\beta = M$
and require $\abs{ \beta \gamma }\ll1$.
The quadratic potential, $-\kappa\,r^2$, embeds the
spherical structure into a curved space at large $r$.

\subsection{Rotation Curves}
\typeout{Rotation Curves}

\begin{figure}
\includegraphics[width=\columnwidth]{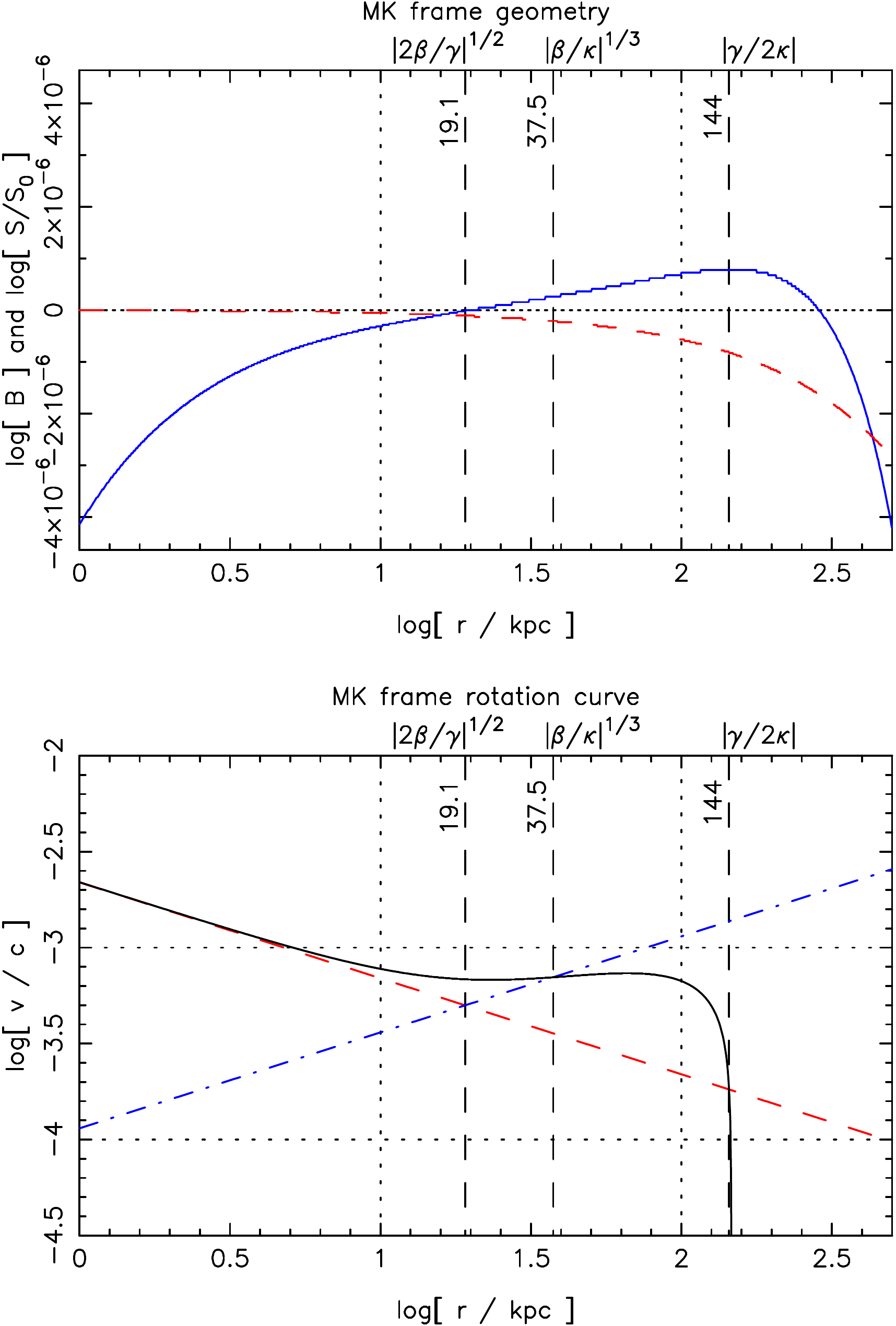}
\caption{ \small
Top panel: The MK metric potential 
	$B(r)=w-2\,\beta/r+\gamma\,r-\kappa\,r^2$
(blue curve) and the associated Higgs halo $S(r)$ (red dashed) 
for a point mass $\beta=M=10^{11}\msun$,
with the MK parameters ($\beta$,$\gamma$,$\kappa$)
used by (Mannheim \& O'Brien~2012)
to fit the rotation curves of spiral galaxies.
Bottom panel: Circular orbit velocity curve 
	$v^2=r\,B^\prime/2\,B$
for the MK potential $B(r)$ (black curve),
	$v^2=\beta/r\,B$
for the Newtonian potential (red dashed) and
	$v^2=\gamma\,r/2\,B$
for the linear potential (blue dot-dash).
Three fiducial radii at transitions
between the Newtonian, linear, and quadratic
potentials are also marked.
The rotation curve is relatively flat
from 10 to 100~kpc as a result of
cancelling contributions from the three potentials.
The potential $B(r)$ has a maximum at the
watershed radius $r=\abs{\gamma/2\,\kappa}\approx144$~kpc,
outside which there are no stable circular orbits.
\label{fig:mk}
}
\end{figure}

 The MK metric's linear potential, $\gamma\,r$, enjoys some successs
in fitting galaxy rotation curves \cite{mo12}.
For circular orbits, the rotation curve for the MK metric is
\begin{equation}
\label{eqn:vv}
	v^2
	\equiv
	\fracd{r^2\,\dot{\theta}^2}{B}
	= \fracd{ \dd \ln{
	\left( \abs{ g_{00} } \right)
	}} {\dd \ln{
	\left( \abs{ g_{\theta\theta} } \right)
	}}
	= \fracd{r\,B^\prime}{2\,B}
	= \fracd{ \fracd{\beta}{r} + \fracd{\gamma}{2}\,r - \kappa\,r^2
	}{ w - \fracd{2\,\beta}{r} + \gamma\, r - \kappa\,r^2 }
\ ,
\end{equation}
where dotted quantities denote time derivatives,
e.g.~$\dot{\theta}=\dd\theta/\dd{t}$.
Fig.~\ref{fig:mk} shows the metric potential $B(r)$ and the
corresponding rotation curve $v(r)$
for a compact point mass $M=\beta=10^{11}\msun$.
In the weak field limit relevant to astrophysics, $B\approx1$,
so that the three terms in the numerator of Eqn.~(\ref{eqn:vv})
determine the shape of the velocity curve.
The Newtonian potential $-2\,\beta/r$ gives a Keplerian
rotation curve $v^2=\beta/r$. The linear potential $\gamma\,r$
gives a rising rotation curve $v^2=\gamma\,r/2$.
A flat rotation curve resembling those
observed on the outskirts of large spiral galaxies
\cite{rft78} then corresponds to the transition between
these regimes, at $r^2\sim2\,\beta/\gamma$,
or $r\sim19$~kpc for the case in Fig.~\ref{fig:mk}.
The velocity at the transition radius,
	$v\approx \left( 2\,\beta\,\gamma\right)^{1/4}$,
can approximate the observed Tully-Fisher relation 
$v^4\propto M$
\cite{tf77}, provided $\gamma$ has the right magnitude
and is independent of $\beta$.
\typeout{1}

\typeout{fig1}

To fit the rising rotation curves observed in smaller dwarf galaxies,
however, 
it is necessary to assume a relationship between $\gamma$ and $M$:
\begin{equation}
        \gamma(M) = \gamma_0
	+ \gamma_\star \left( \fracd{M}{\msun} \right)
	= \gamma_0\, \left( 1 + \fracd{M}{M_0} \right)
\ ,
\end{equation}
where $M$ is the galaxy mass,
$\gamma_0=3.06\times10^{-30}$~cm$^{-1}$,
$\gamma_\star=5.42\times10^{-41}$~cm$^{-1}$,
and $M_0\equiv\gamma_0\,\msun/\gamma_\star=5.6\times10^{10}\msun$.
This metric fits the rotation curves of a wide variety of 
spiral galaxies, replacing their individual dark matter halos
by just 2 free parameters \cite{m93,m97}.

To justify this particular form, it is argued \cite{m93} that 
$\gamma_0$ is generated by matter external to $r$
while $\gamma_\star$ is generated by matter internal to $r$.
This argument is plausible but in our view not really convincing
until it becomes clearer how to 
calculate $\gamma_0$ given the external matter
distribution in the expanding universe.

The most recent fits \cite{mo12} to a sample of 111 galaxies
require invoking the quadratic potential $-\kappa\,r^2$
to counter the rising $\gamma\,r$ potential on the outskirts
of particularly large galaxies like Malin~1.
The required scale, 
$-\kappa=9.54\times10^{-54}$~cm$^{-2}
\sim-(100$~Mpc$)^{-2}$,
is plausibly identified with the observed size of typical
structures in the cosmic web.
For the $10^{11}\msun$ point mass illustrated
in Fig.~\ref{fig:mk},  the transition from rising linear
to falling quadratic potential
has the effect of extending the relatively flat part of the
rotation curve out to around 100~kpc, and 
eliminating bound circular orbits 
outside the ``watershed radius''
$r=\abs{\gamma/2\,\kappa}^{1/2}=144$~kpc,
where the potential $B(r)$ reaches a maximum.

\subsection{ Concerns about using the MK metric}

 Given the notable success of the MK metric in fitting
a wide variety galaxy rotation curves with just 3 parameters,
it is tempting to conclude that CG provides a simpler
description of galaxy dynamics than an alternative
model with hundreds of individual dark matter halos.

 However, the vacuum has a non-zero Higgs field, with
radial profile $S(r)$.
This raises two potential problems with using the MK metric's
linear potential $\gamma\,r$ to fit galaxy rotation curves.
First, with a non-constant $S(r)$, test particles
find their rest masses changing with position,
causing them to deviate from geodesics of the MK metric
\cite{m93b,wm01}.
Second, if $S(r)$ fails to satisfy $\left(1/S\right)^{\prime\prime}=0$,
then $f(r)$ is non-zero, causing the MK coefficients
$w$, $\beta$, $\gamma$, $\kappa$ to be functions of $r$,
and altering their radial dependence within and outside
extended mass distributions such as galaxies and galaxy clusters.

The role of $S(r)$ in sourcing $B(r)$ has been
previously investigated \cite{m07,bv09}, but 
the best CG analysis of galactic rotation curves to date
\cite{mo12} omits this effect, arguing that it is negligible.
In our view conclusions about the success of CG in fitting
galaxy rotation curves are unsafe unless it can be justified
to neglect radial gradients in $S(r)$.
We argue below that even though $S(r)$ is very nearly constant,
its radial gradient is large enough to
significantly alter predictions for galaxy rotation curves,
and moreover the effect on the rotation curve is
to cancel that of the linear potential.

Null geodesics (photon trajectories) 
are independent of conformal transformations,
and those of the MK metric
are well studied \cite{ep98,p04,sk10,vo13}.
A major challenge to CG is that a linear potential 
with $\gamma\,r>0$ is needed to fit galaxy rotation curves,
and this produces light bending in the wrong direction, away from
the central mass rather than toward it, making it 
difficult to account for observed gravitational lensing effects.
However, since $S(r)$ affects $B(r)$, analysis of lensing
by extended sources like galaxies and clusters must also
include the Higgs halo.
We show below that the Higgs halo $S(r)$ outside
a point mass effectively eliminates the $\gamma\,r$ potential,
so that rotation curves may no longer constrain the sign of $\gamma$.
We may then reconsider using $\gamma<0$ when analysing
gravitational lensing effects.


\section{Conformally Coupled Higgs Field}
\label{sec:higgs}
\typeout{Conformally Coupled Higgs Field}

Vacuum solutions of GR, such as the Schwarzschild and Kerr
metrics, assume $T^\mu_\nu=0$, and the MK metric of CG
assumes $f\propto T^0_0-T^r_r=0$.
However, the vacuum now has a Higgs field $S$,
for which $T^\mu_\nu$ and/or $f$ may well not vanish.
A family of analytic solutions of GR with a conformally-coupled
scalar field \cite{wr07} includes the extreme
Reissner-Nordstr\"{o}m black hole metric, with
\begin{equation}
	B(r) = \left( 1 - \fracd{M}{r} \right)^2
\ ,
\end{equation}
sourced by the scalar field profile
\begin{equation}
	S(r) = \left(
	\fracd{3}{4\,\pi} 
	\right)^{1/2} 
	\fracd{M}{r-M}
\ .
\end{equation}
Below we discuss a similar solution for CG.

For the CG matter action
\begin{equation} 
	I_M = \int\,d^4x \sqrt{-g}\, \lag_M
\ ,
\end{equation}
the Lagrangian density \cite{m07} is
\begin{equation}
\label{eqn:lag}
	\lag_M = - \fracd{\sigma}{2} \left(
	S^{;\alpha}\,S_{;\alpha}
	- \fracd{\ricci}{6}\,S^2 \right)
	- \lambda\,S^4
	- \psibar\, \left( \dslash - \muf\,S \right)\, \psi
\ .
\end{equation}
This features a Dirac 4-spinor field $\psi$,
with Dirac operator $\dslash$ and 
Yukawa coupling to the conformal Higgs field $S$
with dimensionless coupling constant $\muf$.
Note that a conformal factor $\Omega$ stretches
the volume element $\sqrt{-g}\,\dd^4x$ by $\Omega^4$,
so that conformal symmetry requires $\lag_M\propto\Omega^{-4}$.
With	$S\propto\Omega^{-1}$,
conformal symmetry holds for 
$S^{;\alpha}\,S_{;\alpha}-\ricci\,S^2/6$,
for the quartic self-coupling potential
	$\lambda\,S^4$,
and as well for the fermion terms with $\psi\propto\Omega^{-3/2}$.
By rescaling $S$ 
the dimensionless parameter $\sigma$ can be set to
$+1$ for a ``right-sign'' or $-1$ for a ``wrong-sign'' 
scalar field kinetic energy.

Varying $I_M$ with respect to $\psi$ gives
the Dirac equation
\begin{equation}
\label{eqn:dirac}
	\dslash\, \psi
	= \muf\, S\, \psi
\ ,
\end{equation}
with the fermion mass mass $m=\muf\,S$ induced by Yukawa coupling.
Varying $I_M$ with respect to $S$ gives
the 2nd-order Higgs equation 
\begin{equation}
\label{eqn:higgs}
	S^{;\alpha}_{;\alpha}
	= \fracd{1}{\sqrt{-g}}\,
	\left( \sqrt{-g}\, g^{\alpha\beta}\,
	S_{,\beta}\right)_{,\alpha}
	= -\fracd{\ricci}{6}\, S
	+ \fracd{4\, \lambda}{\sigma}\, S^3 
	- \fracd{\muf}{\sigma}\,\psibar\,\psi
\ .
\end{equation}
This is the Klein-Gordon equation in curved spacetime,
for a massless scalar field $S$ with 
a fermion source $\muf\,\psibar\,\psi/\sigma$ and
a space-time dependent ``Mexican hat'' potential
\begin{equation}
	V(S) = -\fracd{\ricci}{12}\,S^2
	+ \fracd{\lambda}{\sigma}\,S^4
\ .
\end{equation}
\typeout{mexican hat}

Varying $I_M$ with respect to the metric gives the
conformal stress-energy tensor, with mixed components
\begin{equation}
	T^\mu_\nu
	\equiv \fracd{2}{\sqrt{-g}}
	g^{\mu\alpha}
	\fracd{ \delta I_M }
	{ \delta g^{\alpha\nu} }
	= T^\mu_\nu \left( \psi \right)
	+ \sigma\, T^\mu_\nu \left( S \right)
\ ,
\end{equation}
where
\begin{equation}
\begin{array}{rl}
	T^\mu_\nu \left( S \right)
& \bs\bs
	= \fracd{2}{3} S^{;\mu}\, S_{;\nu}
	- \fracd{1}{3}\, S\, S^{;\mu}_{;\nu}
	- \fracd{1}{6}\, S^2\, \ricci^\mu_\nu
\\ \\ & \bs\bs
	- \delta^\mu_\nu\, \left(
	\fracd{1}{6}\, S^{;\alpha}\, S_{;\alpha}
	- \fracd{1}{3}\, S\, S^{;\alpha}_{;\alpha}
	- \fracd{1}{12}\, \ricci\, S^2
	+ \fracd{\lambda}{\sigma}\, S^4
	\right)
\ .
\end{array}
\end{equation}
The trace
\begin{equation}
	T^\alpha_\alpha
	= \psibar\, \dslash\, \psi
	+ \sigma\, \left(
	S\, S^{;\alpha}_{;\alpha}
	+ \fracd{1}{6}\, \ricci\, S^2
	\right)
	- 4\, \lambda\, S^4
\end{equation}
vanishes by virtue of the 
Dirac and Higgs equations, (\ref{eqn:dirac})
and  (\ref{eqn:higgs}) respectively.

For static spherically-symmetric fermion fields,
the stress-energy tensor takes the form 
\begin{equation}
	T^\mu_\nu\left(\psi\right) ={\rm diag}
	\left(-\rho, p_r, p_\perp, p_\perp \right)
\ ,
\end{equation}
with energy density $\rho$, radial pressure $p_r$,
and azimuthal pressure $p_\perp$.
The Dirac equation (\ref{eqn:dirac}) then gives
\begin{equation}
	T^\alpha_\alpha\left( \psi \right)
	= p_r + 2\, p_\perp - \rho
	= \psibar\, \dslash\, \psi
	= \muf\,S\, \psibar\,\psi
\ .
\end{equation}

 We can consider a Higgs field $S(r,t)$,
allowing for a possible time dependence, with the
understanding that the consequent stress-energy tensor
and gravitational source $f(r)$
must be time independent for the static
spherical structures of primary interest here.
An example is a complex Higgs field with
$S\propto e^{-i\,\omega\,t}$, for which 
$(\dot{S})^2=-\omega^2\,S^2$ and $\ddot{S}=-\omega^2\,S$.

Specialising to the MK metric, the Higgs equation (\ref{eqn:higgs}) 
evaluates as
\begin{equation}
\label{eqn:higgsb}
	\fracd{ \ddot{S} }{B}
	= \fracd{1}{r^2}\left(r^2\,B\,S^\prime\right)^\prime
	+ \fracd{\ricci}{6}\, S
	- \fracd{4\,\lambda}{\sigma}\,S^3
	+ \fracd{ p_r + 2\, p_\perp - \rho }
	{\sigma\,S}
\ ,
\end{equation}
with the Ricci scalar
\begin{equation}
	\ricci = \fracd{
	\left(r^2\,B\right)^{\prime\prime}-2
		}{r^2}
	= \fracd{2\,\left(w-1\right)}{r^2}
	+ \fracd{6\,\gamma}{r}
	- 12\,\kappa
\ .
\end{equation}

\section{ The Higgs Frame }
\typeout{ The Higgs Frame }
\label{sec:frame}

A conformal transformation $\Omega$ maps 
the Higgs field $S$ to
\begin{equation}
	S \rightarrow \tilde{S} = \Omega^{-1}\, S
\ .
\end{equation}
Transforming to the ``Higgs frame", where $\tilde{S}=S_0$,
requires the specific conformal factor $\Omega_S(x)=S(x)/S_0$.
In the static spherical geometry,
given any CG solution $B(r)$ and $S(r)$ in the MK frame,
where $-g_{00} = 1/g_{rr} = B(r)$,
we can ``stretch'' the metric into the Higgs frame:
\begin{equation}
	g_{\mu\nu} \rightarrow
        \tilde{g}_{\mu\nu} 
	=
	\Omega^2\, 
	g_{\mu\nu}
	=
	\left(\fracd{S}{S_0} \right)^2\,	
	g_{\mu\nu}
\ .
\end{equation}
 In the Higgs frame, test particles have space-time 
independent rest masses, $\tilde{m}=\muf\,S_0$,
and thus they follow geodesics of this stretched metric
$\tilde{g}_{\mu\nu}$,
rather than those of the MK metric $g_{\mu\nu}$.

\subsection{Vacuum Stability and Spontaneous Symmetry Breaking}

Using a tilde to denote the Higgs-frame counterparts of the
MK-frame Higgs and fermion fields, we have 
	$\tilde{S}=\Omega^{-1}\,S=S_0$, 
and
$\tilde{\psi}=\Omega^{-3/2}\psi=\left(S/S_0\right)^{-3/2}\,\psi$.
The fermion stress-energy components are then
$\left(\tilde{\rho}, \tilde{p}_r, \tilde{p}_\perp \right)=
\left( \rho, p_r, p_\perp \right)\,(S/S_0)^{-4}$.
The MK-frame Higgs equation is then
\begin{equation}
	\fracd{ \ddot{S} }{B}
	= \fracd{1}{r^2}\left(r^2\,B\,S^\prime\right)^\prime
	+ \fracd{\ricci}{6}\, S
	- 4\, \bar{\lambda}\, S^3
\ ,
\end{equation}
where we define
\begin{equation}
	\bar{\lambda} \equiv 
	\fracd{ \lambda } {\sigma}
	+ \fracd{ \tilde{\rho}
	- \tilde{p}_r
	- 2 \tilde{p}_\perp
	} {4\, \sigma\, S_0^4}
\ .
\end{equation}
Note that the fermions effectively strengthen the
quartic Higgs self-coupling constant.
The corresponding Higgs potential is
\begin{equation}
	V(S) = - \fracd{\ricci}{12}\, S^2
	+ \bar{\lambda}\, S^4
	= \bar{\lambda}\,
	\left( S^2
	- \fracd{\ricci}{24\,\bar{\lambda}} \right)^2
	- \fracd{ \ricci^2 }{ 576\, \bar{\lambda} }
\ .
\end{equation}
A stable vacuum in the MK-frame
requires $\bar{\lambda}>0$, so that
$V(S)$ is bounded from below.
Spontaneous symmetry breaking to 
induce non-zero fermion masses can then occur
for positive curvature $\ricci>0$.
The minimum of $V(S)$ occurs at
 $S^2=\ricci/24\,\bar{\lambda}$.
This gives the vacuum energy density
$V(S) = -\ricci^2/576\,\bar{\lambda} = -\bar{\lambda}\,S^4$.

Note that $\bar{\lambda}>0$ and $\ricci>0$ are
not required, however, since
a time-dependent $S(r,t)$ in the MK frame
corresponds to a constant $S_0$ in the Higgs frame.

\subsection{ Source-Free Solution : The BV Metric }
\typeout{ Analytic Source-Free Solution : The BV Metric }
\label{sec:bv}

The source $f(r)$ in CG's 4-th order Poisson equation
includes
both fermion and Higgs contributions
\cite{m07,bv09}:
\begin{equation}
\label{eqn:f}
        4\,\acg\, f(r) =
	\fracd{3}{B}\, \left( p_r - \rho \right)
	+ \sigma\, S^3\, \left(
	\left( \fracd{1}{S} \right)^{\prime\prime}
	+ \fracd{1}{B^2} \left( \fracd{1}{S} \right)^{\cdot\cdot}
	\right)
\ .
\end{equation}
\typeout{S(r,t)}
The Higgs field $S(r,t)$ makes no explicit contribution to
$f(r)$ if and only if it takes the specific form
\begin{equation}
	S(r,t) = \fracd{ S_0\, t_0\, a }
	{ 
	\left( t + t_0 \right)
	\left( r + a \right)
	}
\ ,
\end{equation}
declining from $S_0$ at time $t=0$ and radius $r=0$
with a timescale $t_0$ and radial length scale $a$.
This holds for either sign $\sigma$.

The time dependence included here may have applications,
for example when embedding static spherical structures
in an expanding universe, with $t_0\sim1/H_0$,
or for a complex Higgs field varying as
$S(r,t)=S(r)\,e^{-i\,\omega\,t}$.
We set $t_0=\infty$ to focus on static solutions.

Note in Eqn.~(\ref{eqn:f})
that a static ``Higgs halo'' $S(r)$ makes a
contribution to $f(r)$ that does not depend on $B(r)$.
Thus when $S(r)$ is known it is straightforward to
integrate the 4th-order Poisson equation (\ref{eqn:poisson})
to determine
the corresponding $B(r)$.  One cannot specify an arbitrary
$S(r)$, however, since $B(r)$ appears in the Higgs
equation (\ref{eqn:higgsb}) for $S(r)$. 
Remarkably, the MK potential $B(r)$ and source-free $S(r)$ 
do admit an analytic solution
\cite{bv09}, as we see below.

 \cite{bv09} (BV) use numerical methods to 
investigate static spherical solutions of CG with
various assumptions about $f(r)$.
Among these BV identify one 3-parameter
analytic solution with a source-free scalar field, 
\begin{equation}
	S(r) = \fracd{S_0\,a}{ r + a }
\ ,
\end{equation}
for which the MK potential is
\begin{equation}
\label{eqn:bvmetric}
	B(r)
	= 
	\left( \fracd{a+r}{a} \right)^2
	\left( 1 - \fracd{ \tilde{h} }{ \tilde{r}(r) } \right)
	- K\,r^2
	\left( 1 - \fracd{ \tilde{h}^3 } { \tilde{r}^3(r) } \right)
\ ,
\end{equation}
\typeout{bv}
where $\tilde{h}\equiv h\,a/(a+h)$,
$\tilde{r}\equiv r\,a/(a+r)$,
and $K \equiv -2\,\lambda S_0^2$.
This metric has a Schwarzschild-like horzon,
with $B(r) \propto (r-h)$ vanishing at $r=h$.

Expanding Eqn.~(\ref{eqn:bvmetric})
in powers of $r$, we can read off
the 3 independent MK parameters $(\beta,\gamma,\kappa)$
in terms of the 3 BV parameters $(h,a,K)$:
\begin{equation}
	2\,\beta = \tilde{h}\, 
	\left( 1 - K\,\tilde{h}^2 \right)
\ ,
\end{equation}
\begin{equation}
	\gamma =
	\fracd{1}{a}
	\left( 2 - 3\,\fracd{\tilde{h}}{a}
	\left( 1 - K\,\tilde{h}^2 \right) \right)
\ ,
\end{equation}
\begin{equation}
	\kappa = K
	- \fracd{1}{a^2}
	\left( 1 - \fracd{\tilde{h}}{a}\,
	\left( 1 - K\,\tilde{h}^2 \right)
	\right)
\ .
\end{equation}
From these one can verify that the constant term,
\begin{equation}
	w = 1 - 
	3\,\fracd{\tilde{h}}{a}
	\left( 1 - K\,\tilde{h}^2 \right)
	= \gamma\,a-1
	= 1 - \fracd{6\,\beta}{a}
\ ,
\end{equation}
satisfies the $W^r_r=0$ constraint
$w^2 = 1 - 6\,\beta\,\gamma$.
The inverse relations are:
\begin{equation}
	a = 
	\fracd{ 1 + w }{\gamma}
	= \fracd{ 6\, \beta }{ 1 - w }
\ ,
\end{equation}
\begin{equation}
	K = \kappa
	+ \left( \fracd{ \gamma }{ 1 + w } \right)^2
	- 2\, \beta
	\left( \fracd{ \gamma }{ 1 + w } \right)^3
\ ,
\end{equation}
and finally, the horizon radius $h$, where $B(r)$ vanishes,
is the smallest positive real root of the cubic
\begin{equation}
	0 = -2\,\beta + w\, r + \gamma\, r^2 - \kappa\, r^3
\ .
\end{equation}

The MK and BV metrics are thus equivalent, 
representing the same 3-parameter source-free solution to the CG 
field equations.
However, as BV show, the Higgs field has a radial profile $S(r)$.
Massive test particles therefore do not follow the time-like
geodesics of the MK metric.

Fortunately, since we know $S(r)$, we know the
conformal transformation between 
the MK frame and the Higgs frame:
\begin{equation}
	S = \fracd{S_0\,a}{r+a} 
	\rightarrow \tilde{S} = \Omega^{-1}\,S = S_0
\ ,
\end{equation}
\begin{equation}
	g_{\mu\nu} \rightarrow 
	\tilde{g}_{\mu\nu}
	= \Omega^2\,g_{\mu\nu}
	= \left( \fracd{S}{S_0} \right)^2\, g_{\mu\nu}
	= \left( \fracd{a}{r+a} \right)^2\, g_{\mu\nu}
\ .
\end{equation}
The stretched metric's circumferential radius
\begin{equation}
	\tilde{r} \equiv
	\sqrt{\abs{\tilde{g}_{\theta\theta}}}
	= \fracd{ r\, S }{ S_0 }
	= \fracd{r\,a}{r+a}
\ 
\end{equation}
maps $0<r<\infty$ to $0<\tilde{r}<a$.
The stretched metric has
\begin{equation}
	\abs{\tilde{g}_{00}} =
	\left(\fracd{a}{r+a} \right)^2\, B(r)
	= 1 - \fracd{2\,M}{\tilde{r}}- K\,\tilde{r}^2
\ ,
\end{equation}
featuring a Newtonian potential with mass
\begin{equation}
	M = \fracd{h}{2}\left( 1 + K\,h^2 \right) 
\ 
\end{equation}
embedded in an external space with curvature $K$.
The conformal transformation does not move the horizon
at $r=h$, which remains at $\tilde{r}=\tilde{h}$.

 Note, however, that while the original MK metric $g_{\mu\nu}$
has a linear potential $\gamma\,r$,
the corresponding Higgs-frame metric $\tilde{g}_{\mu\nu}$
has no term in $\tilde{g}_{00}$ linear in $\tilde{r}$.
Thus even though the Higgs field $S(r)$ declines only
slightly from its central value $S_0$,
this has a significant effect on the shape of the potential
and the resulting rotation curve.
With $B$ rising as $B\approx1+\gamma\,r$,
$S$ falls as $S/S_0\approx1-\gamma\,r/2$,
so that $S^2\,B$ lacks a linear potential.
While this result is demonstrated here for a point mass,
rather than for a more realistic extended source
structure, it indicates the potential
danger when using the linear potential in the
MK metric to fit galaxy rotation curves.

\begin{figure}
\includegraphics[width=\columnwidth]{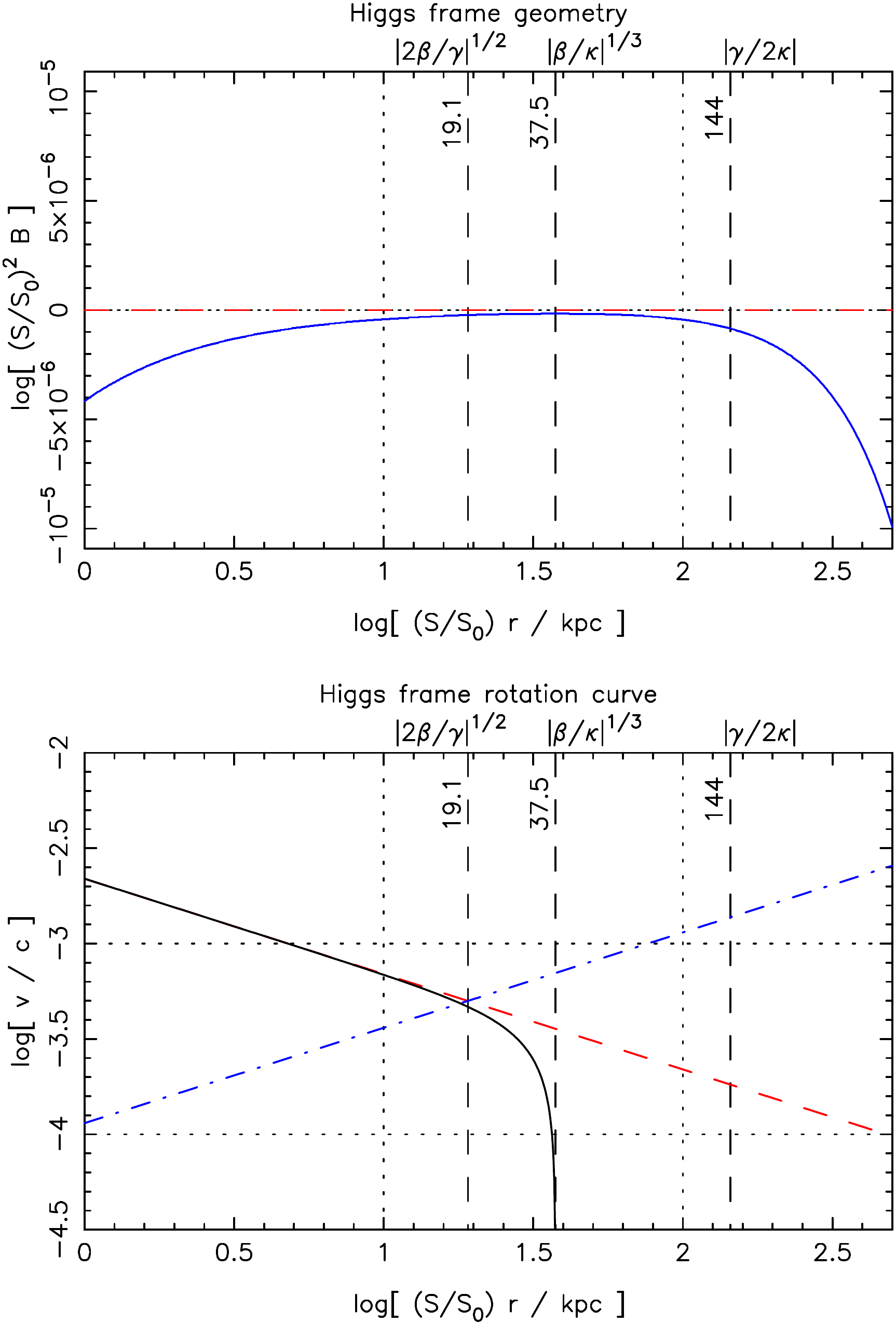}
\caption{ \small
Same as in Fig.~\ref{fig:mk} but after a conformal
transformation $\Omega(r)=S(r)/S_0$ stretches
the geometry from the MK frame, where 
$B(r)=-g_{00}=1/g_{rr}$,
to the Higgs frame, where the Higgs field is constant.
The Higgs-frame potential $\tilde{B}\equiv(S/S_0)^2\,B$
has a maximum at the watershed radius 
$\tilde{r}\equiv (S/S_0)\,r=\abs{\beta/\kappa}^{1/2}\approx37.5$~kpc,
outside which there are no stable circular orbits.
Note that $B(r)$ has a rising linear potential $\gamma\,r$,
but this is effectively canceled by the decline in $S^2$,
so that the rotation curve is Keplerian out to $\sim20$~kpc.
\label{fig:mkh}
}
\end{figure}

Fig.~\ref{fig:mkh} further illustrates this point
by showing the Higgs-frame potential $(S/S_0)^2\,B$,
and the corresponding rotation curve,
for the same $10^{11}\msun$ point mass
as in Fig.~\ref{fig:mk}.
The Higgs field is constant, by definition, in
the Higgs frame. 
Because 
$a\approx2/\gamma=7.6\times10^{10}$~pc
is by far the longest scale in the problem,
$r$ and $\tilde{r}=r\,a/(r+a)$ are nearly identical,
and the Higgs-frame mass $M$ and curvature $K$ are
essentially unchanged from their MK-frame
counterparts $\beta$ and $\kappa$.
The Higgs-frame potential $(S/S_0)^2\,B$ retains
the Newtonian potential $-2\,\beta/\tilde{r}$ and 
the quadratic potential $-\kappa\,\tilde{r}^2$, 
but lacks a linear potential term.
The rotation curve thus follows a Keplerian profile
out to $\sim20$~kpc,
bending down as the quadratic potential takes hold, and
the watershed radius, where $-\tilde{g}_{00}=(S/S_0)^2\,B$ has a maximum, 
is now at $\tilde{r}=\abs{\beta/\kappa}^{1/3}=37.5$~kpc.

\typeout{fig2}

\section{ Astrophysical Tests }
\typeout{ Astrophysical Tests }
\label{sec:eqns}

To really test CG with astrophysical observations is 
considerably harder than simply using geodesics
of the MK metric with $B(r)$ sourced by matter.
The source $f(r)$ in CG's 4th-order Poisson equation
must include contributions from the Higgs halo $S(r)$,
in addition to those from the matter (+radiation)
energy density $\rho(r)$ and pressure $p(r)$.
These are specified in the Higgs frame,
	$\tilde{\rho}(\tilde{r})$
and	$\tilde{p}(\tilde{r})$,
and moved to the MK frame using 
	$\tilde{r}=r\,S(r)/S_0$,
	$\rho(r)=\left(S/S_0\right)^4\tilde{\rho}(\tilde{r})$
and	$p(r)=\left(S/S_0\right)^4\tilde{p}(\tilde{r})$.
For example, to model spherical structures
similar to the matter distribution in galaxies
and clusters, it may be appropriate to
adopt a Hearnquist profile \cite{h90}
\begin{equation}
	\tilde{\rho}(\tilde{r}) = \fracd{\rho_0}{x\left(x+1\right)^3}
\ ,
\end{equation}
with $x=\tilde{r}/r_0$ in units of the scale radius $r_0$,
and with $\rho_0=M/(2\,\pi\,r_0^3)$ 
for total mass $M$.
The enclosed mass profile is
\begin{equation}
	M(\tilde{r}) = M\, \left( \fracd{x}{x+1} \right)^2
\ .
\end{equation}
This Hearnquist profile $\tilde{\rho}(\tilde{r})$ is
specified in the Higgs frame, then scaled by $(S/S_0)^4$
for use in the MK frame where the 
4th-order Poisson equation and 2nd-order Higgs equation 
are more easily solved.

In the MK frame, $B(r)$ and $S(r)$ satisfy
their equations of motion.
The 4th-order Poisson equation for $B(r)$,
\begin{equation}
\begin{array}{rl}
	\fracd{4\,\acg}{r}
	\left( r\,B \right)^{\prime\prime\prime\prime}
& \bs \bs
	= \sigma\,
	S^3\left( \left(\fracd{1}{S}\right)^{\prime\prime}
	+ \fracd{1}{B^2}\left(\fracd{1}{S}\right)^{\cdot\cdot} \right)
\\ & \bs \bs
	- \fracd{3}{B}\,
	\left( \fracd{S}{S_0} \right)^4\,
	\left( \tilde{\rho} + \tilde{p} \right)
\equiv 4\,\acg\, f(r)
\ ,
\end{array}
\end{equation}
is convenient because solutions of the form
\begin{equation}
	B(r) = w(r) - \fracd{2\,\beta(r)}{r}
	+ \gamma(r)\, r - \kappa(r)\, r^2
\end{equation}
can be found for extended sources $f(r)$
by integrating 1st-order equations,
with appropriate boundary conditions:
\begin{equation}
	\beta^\prime = \fracd{r^4}{12}\, f(r)
\ , \hspace{2mm}
	\beta(0)=\beta_0
\ ,
\end{equation}
\begin{equation}
	\gamma^\prime = -\fracd{r^2}{2}\, f(r)
\ , \hspace{2mm}
	\gamma(0)=\gamma_0
\ ,
\end{equation}
\begin{equation}
	\kappa^\prime = -\fracd{r}{6}\, f(r)
\ , \hspace{2mm}
	\kappa(\infty)=\kappa_\infty
\ ,
\end{equation}
\begin{equation}
	w^\prime = \fracd{r^3}{2}\, f(r)
\ , \hspace{2mm}
	w^2(r) = 1 - 6\, \beta(r)\, \gamma(r)
	+\fracd{3\,r^4}{4\,\acg}\,T^r_r(r)
\ .
\end{equation}
The MK parameters, $\beta(r)$, $\gamma(r)$, $\kappa(r)$, $w(r)$,
are then internal and/or external moments of $f(r)$.
For non-singular structures, appropriate boundary conditions
at the origin are $\beta(0)=0$ and 
$\gamma(0)=0$, though non-zero values may also be chosen
for an unresolved
central source such as a nucleon, a star, or a black hole.
The curvature of the external 3-space is set by $\kappa(\infty)$.
The 3rd-order constraint on $w$ can be set any radius
where $T^r_r$ is known.

Note that even for an extended source $f(r)$,
the first 3 derivatives of $B(r)$ evaluate as 
if the MK parameters were $r$-independent.
For example:
\begin{equation}
\begin{array}{rl}
	B^\prime
& \bs\bs
	= \fracd{2\,\beta}{r^2}
	+ \gamma
	- 2\, \kappa\, r
	- \fracd{2\, \beta^\prime}{r}
	+ w^\prime
	+ \gamma^\prime\, r
	- \kappa^\prime\, r^2
\\ \\ & \bs\bs
	= \fracd{2\,\beta}{r^2}
	+ \gamma
	- 2\, \kappa\, r
	+ \left( 
	-\fracd{1}{6}
	+ \fracd{1}{2}
	- \fracd{1}{2}
	+\fracd{1}{6}
	\right)\, r^3\,f
\ .
\end{array}
\end{equation}
As a consequence, the Ricci scalar remains
\begin{equation}
	\ricci
	= \fracd{2\,\left(w(r)-1\right)}{r^2}
	+ \fracd{6\,\gamma(r)}{r}
	- 12\,\kappa(r)
\ ,
\end{equation}
and the 3rd-order constraint remains
\begin{equation}
	W^r_r = \fracd{
	w(r)^2 + 6\,\beta(r)\,\gamma(r)-1
	}{3\,r^4}
	= \fracd{1}{4\,\acg}\,T^r_r
\ .
\end{equation}
Here $T^r_r$ includes radial pressure from both matter (+radiation)
and from the Higgs halo,
\begin{equation}
	T^r_r 
	= \tilde{p}_r\,\left(\fracd{S}{S_0}\right)^4
	+ \sigma\, T^r_r \left( S \right)
\ ,
\end{equation}
with the Higgs halo contribution being,
\begin{equation}
\begin{array}{rl}
	T^r_r \left( S \right)
& \bs \bs
	= \fracd{ \left( \dot{S} \right)^2
	- 2\, S\, \ddot{S} } {6\,B}
	+ \fracd{B}{2}\left(S^\prime\right)^2
	+ \fracd{S\,S^\prime}{6}
	\left( B^\prime
	+ \fracd{4\,B}{r} \right)
\\ & \bs\bs
	+ \fracd{S^2}{6}
	\left( \fracd{B^\prime}{r}
	+ \fracd{B-1}{r^2} \right)
	- \fracd{\lambda}{\sigma}\, S^4
\ .
\end{array}
\end{equation}

Because $f(r)$ depends on $B(r)$ and $S(r)$,
the moment integrals for $B(r)$ must be iterated
along with solving the 2nd-order MK-frame
Higgs equation for $S(r)$:
\begin{equation}
\begin{array}{rl}
	\fracd{\ddot{S}}{B}
	= \fracd{1}{r^2}
	\left( r^2\,B\,S^\prime \right)^\prime
	+ \fracd{\ricci}{6}\, S
	- 4\, \left( \fracd{\lambda}{\sigma}
	+ \fracd{\tilde{\rho}
	- \tilde{p}_r - 2\,\tilde{p}_\perp}
	{4\, \sigma\, S_0^4} \right)\, S^3
\ ,
\end{array}
\end{equation}
with boundary conditions $S(0)=S_0$ and $S^\prime(0)=S_1$.

Having solved for $B(r)$ and $S(r)$ in the MK frame,
we move back to the Higgs frame, and use geodesics of the 
resulting Higgs-frame metric 
$\tilde{g}_{\mu\nu}=(S(r)/S_0)^2\,g_{\mu\nu}$
to test CG in 3 ways:
\\ \hspace{5mm} 1. galaxy rotation curves.
\\ \hspace{5mm} 2. galaxy cluster potentials probed by X-ray gas.
\\ \hspace{5mm} 3. lensing by galaxies and galaxy clusters.

For example, the circular orbit rotation curve is:
\begin{equation}
	v^2 =
	\fracd{\dd\ln{\left(\abs{\tilde{g}_{00}}\right)}}
	{\dd\ln{\left(\abs{\tilde{g}_{\theta\theta}}\right)}}
	= \fracd{\dd\ln{\left( S\,B^{1/2}\right ) }}
	{\dd\ln{\left( S\,r\right)}}
	= \fracd{ v_{B}^2 + v_S^2 }{ 1 +  v_S^2 }
\ ,
\end{equation}
where
\begin{equation}
v_{B}^2 = \fracd{r\,B^\prime}{2\,B}
	= \fracd{
	\fracd{\beta}{r}
	+ \fracd{\gamma}{2}\,r
	- \kappa\,r^2 }
	{ w -\fracd{2\,\beta}{r}
	+ \gamma\,r - \kappa\,r^2 }
\end{equation}
is the rotation curve arising from the 
MK potential $B(r)$, as used by \cite{mo12},
and
	$v_S^2 \equiv r\,S^\prime/S$
implements the corrections arising from the
Higgs halo profile $S(r)$.

 One of the objections to CG is that for the MK metric
with $\gamma>0$ the linear potential causes
light rays to bend away from the point mass,
rather than toward it. But our results show that
a rising MK-frame potential $B(r)$ is compensated
by a corresponding decline in the Higgs halo $S(r)$.
In light of this, the galaxy rotation curves may not
in fact require $\gamma>0$, and we may now reconsider
adjusting the strength and sign of the linear potential
when testing CG predictions for gravitational
bending of light rays.
It remains to be shown whether the effect of an
extended source $f(r)$ appropriate to modelling
galaxies and clusters can fit the light bending angles
from lensing as well as the flat rotation curves.
We hope to address this in future work.

\section{Conclusions}
\typeout{Conclusions}
\label{sec:fini}

The 4th-order field equations of Conformal Gravity
have vacuum solutions that augment the Schwarzschild
metric with linear and quadratic potentials \cite{mk89}.
This MK metric has been used to fit the rotation
curves of a wide variety of galaxies with
only 3 free parameters \cite{mo12}.

We highlight two potential problems with using 
geodesics of the MK metric to study 
rotation curves of galaxies.
First, the MK metric is a source-free solution
to CG's 4th-order Poisson equation, 
but the conformally-coupled Higgs field makes an extended
halo $S(r)$ that also contributes to the gravitational source
unless it has a specific radial profile, $S(r)=S_0\,a/(r+a)$.
Second, since particle masses scale with the Higgs field,
the Higgs halo $S(r)$ pushes test particles off
geodesics of the MK metric.

To address these issues, we note that
a conformal factor $\Omega(r)=S(r)/S_0$ stretches the metric
to a form that makes the Higgs field constant.
Test particles then follow geodesics of this
stretched ``Higgs-frame" metric.

For the analytic solution to the source-free CG equations
\cite{bv09}, which is equivalent to the MK metric,
we find that the effect of stretching the metric to
the Higgs frame is to eliminate the
linear potential that is used to fit galaxy rotation curves.
Thus the remarkable results of \cite{mo12},
using geodesics of the MK metric to fit a large variety
of galaxy rotation curves with just 3 parameters, 
may be testing an empirical model rather than
the actual CG predictions.

We collect the equations and outline the procedure
for astrophysical tests of CG in static spherical geometries.
Specifically, the sources for CG's 4th-order Poisson equation
include not only the energy density and pressure
of distributed matter (stars+gas),
and radiation if relevant,
but also the associated Higgs halo $S(r)$.
The resulting MK metric $g_{\mu\nu}$
must then be ``stretched'' to the Higgs frame metric 
$\tilde{g}_{\mu\nu}=(S/S_0)^2\,g_{\mu\nu}$.
The Higgs-frame geodesics then provide predictions
for testing CG against observations.


\subsection*{Acknowledgements}

KH would like to thank Amy Deacon, Aidan Farrell
and Indar Ramnarine for
hospitality at the University of the West Indies in Trinidad,
Alistair Hodson, Alasdair Leggat, and Carl Roberts
for helpful comments on the presentation,
an anonymous referee for thought provoking criticism,
and the UK Science and Technology Facilities
Council (STFC) for financial support through
consolidated grant ST/M001296/1.

\bsp	
\label{lastpage}
\end{document}